# A Lagrangian Relaxation for the Maximum Stable Set Problem[*]

Manoel Campêlo[†]
Departamento de Estatística e Matemática Aplicada
Universidade Federal do Ceará, Brazil

Ricardo C. Corrêa[†]
Departamento de Computação
Universidade Federal do Ceará, Brazil

November 11, 2018

## Abstract

We propose a new integer programming formulation for the problem of finding a maximum stable set of a graph based on representatives of stable sets. In addition, we investigate exact solutions provided by a Lagrangian decomposition of this formulation in which only one constraint is relaxed. Some computational experiments were carried out with an effective multi-threaded implementation of our algorithm in a multi-core system, and their results are presented.

## 1 Introduction

With the emergence of multicore computers, methods of parallelizing performance-critical applications, particularly combinatorial optimization problems, become a relevant issue. One major question is the identification of algorithmic methods amenable to parallel treatment in a shared memory setting. In this context, decomposition appears as a useful tool to benefit from the potential of the processing power made available by many cores. The objective of this paper is to show a simple and effective decomposition method for a classical combinatorial problem and its implementation in a multicore system.

### 1.1 The problem and its mathematical formulation

Let $G = (V, E)$ be an undirected, simple, nonempty and connected graph. We write $n$ and $m$ for $|V|$ and $|E|$, respectively. A *stable set* of $G$ is a subset $W \subseteq V$ such that, for every pair $u$ and $v$ of distinct vertices in $W$, $uv \notin E$, where $uv$ is the notation for the edge of $G$ defined by $u$ and $v$. Stable sets induce structures in graphs that model constraints over unrelated elements of a given set (represented by $V$) when some pairs of elements are in conflict (represented by $E$). In such situations, $G$ can be seen as representing logical relations between elements such that every edge $uv$ defines a valid inequality of the form

$$x_u + x_v \leq 1, \qquad (1)$$

being $x_u$ and $x_v$ binary variables associated with $u$ and $v$, respectively. In other words, only vertices of a stable set of $G$ are allowed to get 1 in a feasible solution [3].

[*]This work is partially supported by CNPq and PADCT projects. Parts of it were done when R. Corrêa was visiting the Universidade Federal do Rio de Janeiro, COPPE/Programa de Engenharia de Sistemas e Computação, RJ, Brazil, and M. Campêlo was visiting the Tepper School of Business, Carnegie Mellon University, USA.

[†]Partially supported by the Conselho Nacional de Desenvolvimento Científico e Tecnológico, CNPq, Brazil. Addresses: Campus do Pici, Bloco 910, 60455-760 Fortaleza, CE, Brazil, {`mcampelo, correa`}`@lia.ufc.br`

The canonical problem involving stable sets is related to the stability number of graphs. Given that $\mathcal{W}$ denotes the family of all stable sets of $G$, the *stability number* $\alpha(G)$ of the graph $G$ is the maximum cardinality of a member of $\mathcal{W}$ [18]. The *Stable Set Problem (SSP)* consists of determining $\alpha(G)$, for a given (arbitrary) graph $G$. A number of structural results and algorithmic techniques exist in the literature for this problem (see, for example, [1] for its properties related to perfect graphs, [4, 11, 12, 15] for polyhedral properties, [3, 9] for its use for generating valid inequalities in general mixed integer problems, [22] for its use in a branch-and-cut algorithm, and [6, 19] for lift-and-project relaxations). The formulation usually adopted in these studies is

$$\alpha(G) = \max\{x(V) \mid x \in STAB(G)\}, \qquad (2)$$

where $x$ is a size-$n$ binary vector indexed by the vertices of $G$, $x(U)$, for $U \subseteq V$, stands for $\sum_{u \in U} x[u]$ and $STAB(G)$ is the convex hull of the incidence vectors of stable sets (including the empty stable set) of $G$, also referred to as the *stable set polytope of* $G$. In spite of the fact that unstructured SSPs are difficult integer programming problems (indeed, it is NP-Hard to approximate $\alpha(G)$ within a factor $n^{\varepsilon}$ for some $\varepsilon > 0$ [2]), the main aspect to be observed in connection with (2) is its ability to provide tight upper bounds when a partial description of $STAB(G)$ is used. In these terms, the number of variables employed, in $O(n)$, is clearly an advantage. However, the number of constraints is a drawback that can be partially circumvented with the cutting-plane method for inequalities of the type

$$x(U) \leq \alpha(G[U]), \text{ where } U \subseteq V \text{ and } G[U] \text{ stands for the subgraph of } G \text{ induced by } U,$$

referred to as *rank inequalities* [4, 22, 25].

## 1.2 Enumeration

There are two enumeration approaches usually adopted to solve the SSP (or the equivalent problem of finding a maximum stable set in the complement of $G$, known as the *Maximum Clique Problem*) to optimality. In both of them, a search tree of subproblems is traversed. The first approach is the best-first branch-and-cut where the cutting-plane method is applied to determine upper bounds for the subproblems and the traversal of the search tree is conducted according to a non-increasing order of such bounds [22]. We concentrate our study in this approach motivated by two main facts. Firstly, best-first is well suitable for parallel treatment; secondly, since several other problems can be formulated as stable set problems, polyhedral and cutting techniques tend to have broader applicability [3, 9, 17].

The second enumeration approach is the combinatorial-based depth-first branch-and-bound obtained by introducing a pruning strategy in a backtrack enumeration. Computational experiments reported in the literature show that this approach can be efficiently implemented, with good performance when dense graphs are considered. To achieve this good performance in practice, the bounds used to guide the traversal of the tree are obtained with combinatorial greedy heuristics than run faster than polyhedral techniques [13, 21, 23]. The accuracy of such bounds highly depends on an ordering of the vertices.

## 1.3 The reformulation and the decomposition

The decomposition we propose for the SSP is based on a reformulation of the problem by means of the notion of representatives of stable sets. The use of representative vertices is an approach that appeared originally in the literature to formulate integer coloring problems [7, 8]. It has been used with other stable set problems successfully [10, 14]. It consists of partitioning the family of stable sets of $G$ into $n$ sub-families, each of them having a vertex of $G$ as a representative. This simple technique allows the consideration of disjoint collections of stable sets almost independently.

In this paper, we propose the use of representative vertices also to formulating the SSP in a way that leads to a natural decomposition of (2) into at most $n$ subproblems, each one being an instance of the SSP associated with an induced subgraph of $G$. The SSP becomes the problem of finding the largest stable set represented by each vertex and, among these, taking the largest one. This decomposition potentially improves the performance of the cutting-plane method since it allows the application of separation heuristics to graphs that are smaller than $G$, at the cost of increasing the number of variables by a factor that may be as large as $n$.

The most important property of the reformulation is that the SSPs derived for each representative are loosely coupled, which makes it simple to attain our objective of decomposing the original problem into smaller independent subproblems. A Lagrangian relaxation is then described that incorporates a single constraint in the objective function. This technique appeared originally in [16] for more unstructured problems. In the case studied in this paper, the decomposition of the relaxed problem becomes much simpler due to the structure of the representatives formulation. As a result, we are left with a number of smaller independent SSP subproblems, each of them associated with a certain subgraph of $G$. Some computational experiments are presented to show that this decomposition can be solved at reasonable computational cost for medium size graphs. In addition, results obtained with a multi-threaded implementation are reported, showing that the decomposition we propose is suitable to be naturally parallelized and executed in nowadays multi-core architectures.

## 1.4 Organization of the paper

The paper is organized as follows. The notation adopted is stated in Section 2. In Section 3, we present the representatives formulations of the SSP and an algorithm to solve the reduced subproblems generated by the Lagrangian decomposition. Finally, we close the paper with the experimental results (Section 4) and conclusions (Section 5).

# 2 Preliminaries

## 2.1 Notation

The complementary graph of $G$ is denoted $\bar{G} = (V, \bar{E})$, and the number of edges $|\bar{E}|$ is represented by $\bar{m}$. A *clique* of $G$ is a stable set of $\bar{G}$. A *clique cover* is a partition of $V$ into cliques of $G$. In the definitions in the sequel, $v$ is a member of $V$. A vertex $u \in V$, $u \neq v$, is called a $v$'s *neighbor* if $uv \in E$. Otherwise, $u$ and $v$ are *anti-neighbors*. The notation $N(v)$ stands for the *neighborhood of $v$* in $G$, i.e. the set $\{u \in V \mid uv \in E\}$. On the other hand, the neighborhood of $v$ in $\bar{G}$, called the *anti-neighborhood of $v$*, is denoted by $\bar{N}(v)$.

Let $W$ be a subset of $V$. Write $E[W]$ for the set of edges of $G[W]$, which in turn is the subgraph of $G$ induced by $W$. In the particular case when $W$ is $\bar{N}(v)$, then we use $G(v)$ for $G[\bar{N}(v)]$.

An *orientation* of $G$ is a mapping $\sigma : E \to V$ such that $\sigma(uv) \in \{u, v\}$. Define the *out-neighborhood* of $u$ as $N^+(u) = \{v \in N(u) \mid \sigma(uv) = v\}$ and its *in-neighborhood* to be $N^-(u) = \{v \in N(u) \mid \sigma(uv) = u\}$. A vertex $s \in V$ is a *source in $\sigma$* if $\sigma(sv) = v$, for all $v \in N(s)$. The orientation $\sigma$ is *acyclic* if every clique has exactly one source in $\sigma$. If an orientation is given for $\bar{G}$, the out- and in-anti-neighborhoods can be defined similarly and be denoted by $\bar{N}^+(u)$ and $\bar{N}^-(u)$, respectively. Additionally, we write $G^-(v)$ instead of $G[\bar{N}^-(v)]$ and $G^+(v)$ for $G[\bar{N}^+(v)]$.

## 2.2 Scaling $STAB(G)$

The $n$-dimensional *characteristic vector* $x_W$ of a stable set $W \in \mathcal{W}$ is such that $x_W[u]$ gets 1 if $u \in W$, and 0 if $u \in V \setminus W$. Let $\mathcal{W}$ also denote the set constituted by the characteristic vectors of stable sets of $G$, i.e. by all vectors $x \in \{0,1\}^n$ such that every *edge inequality*

$$x[u] + x[v] \leq 1, \quad uv \in E,$$

holds. $STAB(G)$ is then given by the convex hull of vectors in $\mathcal{W}$. Thus, each vector $x \in STAB(G)$ is a convex combination of the form

$$x = \sum_{W \in \mathcal{W}} \lambda_W x_W, \quad \sum_{W \in \mathcal{W}} \lambda_W = 1, \lambda_W \in [0,1], \text{for all } W \in \mathcal{W}.$$

The version of a set $P$ scaled by $p \in \mathbb{R}$ is obtained by multiplying the vectors of $P$ by $p$ (multiplying a vector by a scalar corresponds to multiplying every entry of the vector by this scalar). In this sense, the polytope $STAB(p, G)$, obtained by scaling $STAB(G)$ by $p$, is the convex hull of vectors in $\mathcal{W}$ scaled by $p$. Therefore, a vector $x \in STAB(p, G)$ can be expressed as

$$x = \sum_{W \in \mathcal{W}} \lambda_W x_W, \sum_{W \in \mathcal{W}} \lambda_W = p, \lambda_W \in [0, p], \text{ for all } W \in \mathcal{W}. \tag{3}$$

It follows that every vector $x \in \{0, p\}^n$ of $STAB(p, G)$ satisfies the *edge inequality, scaled by p*, i.e.

$$x[u] + x[v] \leq p, \quad uv \in E.$$

Special cases of scaled stable set polytopes are $STAB(0, G) = \{\mathbf{0}\}$, where $\mathbf{0}$ denotes the null vector, and $STAB(G) = STAB(1, G)$, which gives

$$\alpha(G) = \text{Max } x(V), \text{ subject to } x \in STAB(1, G).$$

## 3 Problem reformulation and its reduction

Assume we are given an acyclic orientation $\sigma$ of $\bar{G}$ and a total order $\prec$ on $V$ that respects $\sigma$, which means that $\sigma(uv) = v$ if, and only if, $u \prec v$, for every $uv \in \bar{E}$. Let the in- and out-anti-neighborhoods be defined according to $\sigma$.

### 3.1 Representatives formulation

A way to describe a family $\{W_1, \ldots, W_k\}$ of $k$ pairwise disjoint stable sets of $G$ is to take $r_i$, the minimal vertex of $W_i$ according to $\prec$, to be the representative of $W_i$ that represents $W_i \setminus \{r_i\} \subseteq \bar{N}^+(r_i)$, for all $i \in \{1, \ldots, k\}$. From this point of view, the family of stable sets is seen as a set of representatives $\{r_1, \ldots, r_k\}$ and, for each representative $r_i$, a set of vertices (maybe empty) represented by $r_i$. Observe that $W_i$ is the result of the union of its $r_i$ with a stable set of $G^+(r_i)$. This fact gives raise to a reformulation of the SSP, as follows. Define a binary variable $x_u$, for all $u \in V$, to indicate whether $u$ is a representative, and a vector $x^u$ of binary variables indexed by the vertices in $\bar{N}^+(u)$ to indicate the vertices represented by $u$. With these variables, we write

$$\alpha(G) = \text{Max} \sum_{u \in V \setminus T} x_u + x^u(\bar{N}^+(u)) \tag{4}$$

$$\text{Subject to } x^u \in STAB(x_u, G^+(u)), \quad u \in V \setminus T \tag{5}$$

$$\sum_{u \in V \setminus T} x_u \leq 1 \tag{6}$$

$$x_u \in \{0, 1\}, \quad u \in V \setminus T \tag{7}$$

The constraints (6)–(7) and the fact that $STAB(0, G) = \{\mathbf{0}\}$ assure that every non-null feasible solution $\hat{x}_u$, $\hat{x}^u$, for all $u \in V \setminus T$, contains exactly one representative, say $r$. In mathematical terms, $\hat{x}_r = 1$ and $\hat{x}_u = 0$, for all $u \in V \setminus T$, $u \neq r$. This leads all the scaled stable set polytopes in (5), except that of $r$, to be inactive, which means that they only contain the null vector. Since the scaled stable set polytopes are disjoint and the objective function is linear, the (non-null) optimal solution will be attained at a vertex of the unique active polytope $STAB(x_r, G^+(r))$, that is, at a characteristic vector of a stable set of $G^+(r)$. The objective function chooses $r$ so that the stability number of $G^+(r)$ is maximum.

The set $T \subseteq V$ is composed by vertices that cannot represent a stable set larger than those represented by the vertices in $V \setminus T$. Such a set generalizes the notion of simplicial vertices used to fix variables in [20]. It can be obtained iteratively as follows. Given an edge $uv$ in $E[V \setminus T]$, the vertex $v$ can be included in $T$ if $\bar{N}^+(v) \subseteq \bar{N}^+(u)$. This condition implies that if $W$ is a stable set represented by $v$, then $W - v + u$ is a stable set represented by $u$.

## 3.2 Lagrangian relaxation and its decomposition

A simple inspection of the representatives formulation shows that it comprises disjoint SSP subproblems coupled by inequality (6). Thus, the SSPs become independent by simply bringing the coupling inequality in the objective function with a penalty $\lambda \in \mathbb{R}_+$. Defining this Lagrangian relaxation of the representatives formulation, we get the Lagrangian function

$$L(x, \lambda) = \sum_{u \in V \setminus T} \left( x_u + x^u(\bar{N}^+(u)) \right) + \lambda(1 - \sum_{u \in V \setminus T} x_u),$$

for $\lambda \geq 0$ [5]. The resulting separable Lagrangian problem is given by

$$\alpha_\lambda = \lambda + \sum_{u \in V \setminus T} \alpha_\lambda^u,$$

where

$$\alpha_\lambda^u = \max \left\{ (1 - \lambda)x_u + x^u(\bar{N}^+(u)) : x^u \in STAB(x_u, G^+(u)), x_u \in \{0, 1\} \right\}. \tag{8}$$

Note that $\alpha_\lambda^u = \max\{\alpha(G^+(u)) + 1 - \lambda, 0\}$, for every $u \in V \setminus T$. This decomposition of the problem in subproblems, each of them being an instance of the SSP in a subgraph of $G$, can be solved with a combination of the Lagrangian dual descent method to compute $\lambda$ with the branch-and-cut method to solve each subproblem, as described next.

## 3.3 Solving the dual Lagrangian problem

In Lagrangian approaches, it is usual to define a dual problem, which consists of finding a configuration of the multipliers that optimize the bound obtained with the relaxation. For any $\lambda \geq 0$, it is known that $\alpha_\lambda$ is an upper bound for $\alpha(G)$ [5]. In particular, if the Lagrangian multiplier $\lambda$ is $\alpha(G)$, then $\alpha_\lambda^u = 0$, for all $u \in V \setminus T$, and therefore $\alpha_\lambda = \alpha(G)$. It follows that an optimum value of $\lambda$ is precisely $\alpha(G)$. Moreover, the maximum value of $L(x, \alpha(G))$ is attained when $x$ is such that $\sum_{u \in V} x_u = 1$.

To compute the optimum multiplier, we use an iterative algorithm. Starting with an initial estimate $\lambda^0$, our algorithm generates, at iteration $\ell \geq 0$, a new estimate $\lambda^{\ell+1}$ depending upon the solution of a relaxation of a subproblem. The iterations are described in Algorithm 1, where the subproblem $LP_u$ mentioned at line 7 is a linear relaxation of (8) (we give more details of $LP_u$ in the next section). Unsolved subproblems are ordered in a non-increasing order of upper bounds in the priority queue $Q$. The branch-and-cut tree of each subproblem is traversed only once in a best-first order (lines 12–24) and the Lagrangian multiplier is updated simultaneously with the traversal of the trees. At iteration $\ell$, the second largest upper bound $\bar{\alpha}'$ of the subproblems in $Q$ gives the estimate $\lambda^\ell$ as indicated in line 13, which is used to guide the search in the sense that all nodes in a selected search tree whose values are greater than $\lambda^\ell$ are explored (lines 17–24). This means that the branch-and-cut algorithm proceeds with a subproblem as long as the associated upper bound is large enough.

An interesting characteristic of Algorithm 1 is that the estimate of the Lagrangian multiplier decreases monotonically until a large stable set is found at line 22. This starts a phase where the multiplier increases monotonically until its optimum value is attained. A variable $x_u$ is fixed at zero when the maximum stable set that can be represented by $u \in V$ is at most as large as the largest stable set found so far.

Since the subproblems in $Q$ are independent, line 18 is performed in parallel on different subproblems.

# 4 Computational experiments

Here we present the results of computational experiments carried out with the representatives formulation and the Lagrangian decomposition. In what follows, we briefly describe some details of the implementations and its comparison with the standard model based on the $STAB(G)$ polytope (formulation (2)). One main interest of this comparison is the investigation of the tradeoff between the increase of the number of variables

**Algorithm 1** Solving the Lagrangian relaxation

1: Define a priority queue $Q$ of subproblems, ordered according their upper bounds
2: $\underline{\alpha} \leftarrow 0$
3: **for all** $u \in V$ **do**
4:     $\underline{\alpha} \leftarrow \max\{\underline{\alpha}, 1 + \text{ size of a maximal stable set of } G^+(u)\}$
5:     $\bar{\alpha}_u \leftarrow 1 + \text{ size of a clique cover of } G^+(u)$
6:     **if** $\bar{\alpha}_u > \underline{\alpha}$ **then**
7:         Generate the linear program associated with $G^+(u)$ and call it subproblem $LP_u$
8:         Add $LP_u$ to $Q$, associated with its current upper bound $\bar{\alpha}_u$
9:
10: $\lambda \leftarrow$ second maximum upper bound of the subproblems generated above
11: $Q \leftarrow \emptyset$
12: **while** $Q \neq \emptyset$ **do**
13:     $\lambda \leftarrow \left\lfloor \bar{\alpha}' - \pi \frac{\bar{\alpha}' - \underline{\alpha}}{|Q|} \right\rfloor$
14:     $Q' \leftarrow \{u \in Q : \lambda \leq \bar{\alpha}_u\}$
15:     **for all** $u \in Q'$ **do in parallel**
16:         Remove subproblem $LP_u$ from $Q$
17:         **while** $LP_u$ is unsolved, $\bar{\alpha}_u > \underline{\alpha}$, and $\lambda \leq \bar{\alpha}_u$ **do**
18:             Select an unsolved node of $LP_u$ and apply the cutting-plane method to it
19:             Update $\bar{\alpha}_u$ according to the upper bounds of the remaining unsolved nodes of $LP_u$
20:             **if** best known integral solution of $LP_u$ is greater than $\lambda$ **then**
21:                 $\underline{\alpha} \leftarrow$ value of the new solution
22:                 Remove from $Q$ all nodes with upper bound at most $\underline{\alpha}$
23:             **if** $\bar{\alpha}_u > \underline{\alpha}$ **then**
24:                 Perform a branching of the branch-and-cut method associated with $LP_u$
25:
26:     **for all** $u \in Q'$ **do in parallel**
27:         **if** $LP_u$ is unsolved and $\bar{\alpha}_u > \underline{\alpha}$ **then**
28:             Add $LP_u$ to $Q$, associated with its current upper bound $\bar{\alpha}_u$
29:
30:     $\pi \leftarrow \max\{\pi/2, 0.05\}$

and the strengthening of the subproblems obtained by the decomposition. The second aspect investigated with the experiments is the performance of the parallel versions of the decomposition.

In all implementations, the data structure used to store the graph is a 0-1 adjacency matrix. The implementation of formulation 2 consists of the construction of an initial linear relaxation and a standard branch-and-cut solver. To construct the initial linear relaxation, every edge of the graph generates an inequality of type (1). Moreover, a greedy algorithm is used to determine a clique cover of $G$. A rank inequality associated with each clique of such cover is added to the linear relaxation.

Our implementations of (4)–(7) consists of the optimization of a core linear program (referred simply to as $LP_u$ in Algorithm 1) containing a partial description of $STAB(G^+(u))$, for each $u \in V \setminus T$. Further details of these implementations are given next.

## 4.1 Pre-processing

Before solving the formulations, our algorithm performs a pre-processing computation on $G$. One task accomplished by the pre-processing is the determination of an orientation of $\bar{G}$, which is used to define the variables of the formulations. The orientation is based on an ordering of the vertices obtained with the following iterative algorithm. At each iteration $\ell = 0, 1, \ldots, n-1$, the $\ell$th vertex in the order is determined. Let $V^\ell$ denote the set of vertices that remain to be ordered after $\ell$ iterations. Naturally, $V^0 = V$. The vertex $u$ to order at iteration $\ell$ is chosen in such a way that the number of edges connecting $u$ to vertices in $V^\ell$ is maximized. To break ties, the strategy proposed in [23] is used. In this strategy, a vertex whose anti-neighborhood induces a denser subgraph is ordered last. According to these criteria, the very first vertex of the ordering is one among those with maximum degree and, in addition, a maximal stable set is placed at the last positions of the order. In Table 1, some properties of the ordering obtained are shown.

The second task of the pre-processing is to determine bounds for $\alpha(G)$, which is done based on the order on the vertices. We take as a lower bound the size of the maximal stable set defined by the last vertices in the order. This is accomplished by the ordering algorithm, without any increase in the time complexity. The lower bounds so obtained are shown in the last column of Table 1. As an upper bound for $\alpha(G)$, a simple greedy heuristic, guided by the order on the vertices, is used to determine a clique cover of $G^+(u)$. This clique cover also gives upper bounds for the subproblems. For each $u \in V \setminus T$, we simply take the number of cliques intersecting $\bar{N}^+(u)$ as the initial upper bound indicated at line 5 of Algorithm 1. The comparison of these upper bounds with the lower bound leads the pre-processing phase to discard those subproblems that do not contain any larger stable set. In Table 1, the number of valid vertices corresponds to the number of subproblems that are not discarded. There are two measures to evaluate the density of the graphs associated with the subproblems, namely: their maximum and average values.

## 4.2 Handling the core linear programs

The construction of the core linear program of a subproblem is also based on the order on the vertices, which gives the definition of the variables and initial constraints. The clique cover of $G$ (that determining the upper bound for $\alpha(G)$) is used to generate rank inequalities, added to the core linear program together with the corresponding edge inequalities. It should be noticed that this order is different from that one used to produce the clique cover used in the construction of the linear program associated with the implementation of the formulation derived from $STAB(G)$.

Each core linear program evolves from its initial state driven by a branch-and-cut algorithm, which has already been proved to be effective to this problem [17, 22]. The separation heuristic at line 18 in Algorithm 1 is implemented in the linear program solver used, CPLEX 11.0. The parameters (including the choice of separation heuristics to apply) used to obtain the results are determined using the tune facility of CPLEX 11.0 with the problem defined on the $STAB(G)$. These are the parameters used for solving the Lagrangian decomposition. The times reported next do not consider the time required to initialize the core linear programs.

|  | Instance |  | Ordering properties | | | |
| --- | --- | --- | --- | --- | --- | --- |
| $G$ | $n$ (density) | $\alpha(G)$ | Valid vertices | Max subgraph density | Av. subgraph density | Stable set size |
| Random graphs | | | | | | |
| g200.90 | 200 (0.90) | 4.6 | 168.0 | 0.979 | 0.855 | 3.6 |
| g200.70 | 200 (0.70) | 7.2 | 186.4 | 0.720 | 0.616 | 5.4 |
| g200.50 | 200 (0.50) | 11 | 187.8 | 0.507 | 0.407 | 8.2 |
| g200.30 | 200 (0.30) | 18 | 183.6 | 0.302 | 0.221 | 14.2 |
| DIMACS graphs | | | | | | |
| c-fat200.1 | 200 (0.92) | 12 | 0 | – | – | 12 |
| c-fat200.2 | 200 (0.84) | 24 | 2 | 0.253 | 0.248 | 24 |
| c-fat200.5 | 200 (0.57) | 58 | 7 | 0.268 | 0.249 | 58 |
| c-fat500.1 | 500 (0.93) | 14 | 1 | 0.265 | 0.265 | 14 |
| c-fat500.2 | 500 (0.63) | 26 | 1 | 0.242 | 0.242 | 26 |
| c-fat500.5 | 500 (0.96) | 64 | 11 | 0.243 | 0.234 | 64 |
| c-fat500.10 | 500 (0.81) | 126 | 28 | 0.245 | 0.234 | 126 |
| brock200.2 | 200 (0.50) | 12 | 187 | 0.512 | 0.416 | 7 |
| brock200.4 | 200 (0.44) | 17 | 186 | 0.347 | 0.258 | 12 |
| brock400.2 | 400 (0.25) | 29 | 372 | 0.255 | 0.192 | 19 |
| brock400.4 | 400 (0.25) | 33 | 379 | 0.252 | 0.192 | 19 |
| p_hat300-1 | 300 (0.76) | 8 | 282 | 0.720 | 0.591 | 8 |
| p_hat300-2 | 300 (0.51) | 25 | 273 | 0.423 | 0.241 | 24 |
| p_hat300-3 | 300 (0.26) | 36 | 266 | 0.243 | 0.132 | 31 |
| hamming8.4 | 256 (0.36) | 16 | 240 | 1.000 | 0.582 | 1 |
| keller4 | 171 (0.35) | 11 | 154 | 0.383 | 0.387 | 8 |
| san200.0.7.2 | 200 (0.30) | 18 | 198 | 1.000 | 0.145 | 1 |
| san200.0.9.1 | 200 (0.10) | 70 | 186 | 0.105 | 0.035 | 12 |
| san200.0.9.2 | 200 (0.10) | 60 | 186 | 0.099 | 0.045 | 12 |
| C125.9 | 125 (0.10) | 34 | 90 | 0.101 | 0.058 | 33 |
| C250.9 | 250 (0.10) | [44,48] | 211 | 0.101 | 0.063 | 37 |
| gen200.9.44 | 200 (0.10) | 44 | 166 | 0.100 | 0.057 | 30 |
| gen200.9.55 | 200 (0.10) | 55 | 159 | 0.099 | 0.059 | 36 |
| mann.a27 | 378 (0.01) | 126 | 375 | 1.000 | 0.033 | 1 |
| mann.a45 | 1035 (0.004) | 345 | 1032 | 1.000 | 0.014 | 1 |

Table 1: Properties of the graphs associated with the subproblems

## 4.3 Results

The algorithms have been implemented in Java and C on a dual quad-core based computer, with 4GBytes of memory and 3.0GHz of clock frequency. We used as test-bench some uniform random graphs and some instances extracted from DIMACS implementation challenge [24]. These instances are organized in Table 2 according to the density of the graphs since it is the key parameter in the comparisons performed. The results reported are the following: the size of the graph (number of vertices $n$ and density), the value of an optimum solution and the initial lower bound, and information about the two implemented approaches. For the implementation of formulation (2), the upper bound given by the fractional solution of the initial core linear program, the total number of branch-and-cut nodes explored, and the total running time (in seconds) for solving the core linear program. For the implementation of Algorithm 1, besides the total number of branch-and-cut nodes explored by all core linear programs and the corresponding running times, the initial value $\lambda^0$ of Lagrangian multiplier is also provided. The values shown for each random graph are the averages over 5 instances. The results for the parallel (multi-threaded) implementation were obtained using four concurrent threads.

| | Instance | | $STAB(G)$ | | | Lagrangian | | |
|---|---|---|---|---|---|---|---|---|
| $G$ | $n$ (density) | $\alpha(G)$ (initial $\underline{\alpha}$) | LP | Nodes | Time | $\lambda^0$ | Nodes | Time |
| Random graphs | | | | | | | | |
| g200.90 | 200 (0.90) | 4.6 (3.4) | 10.36 | 428.60 | 9.92 | 13.4 | **400.20** | **4.99** |
| g200.70 | 200 (0.70) | 7.2 (5.4) | 21.53 | 2645.60 | 100.35 | 24.6 | **1251.2** | **25.76** |
| g200.50 | 200 (0.50) | 11 (8.2) | 24.70 | 15098.80 | 208.38 | 36.8 | **11922.0** | **64.02** |
| g200.30 | 200 (0.30) | 18 (14) | 35.65 | 105922.4 | 879.05 | 52.4 | **112019.4** | **428.86** |
| DIMACS graphs | | | | | | | | |
| c-fat200.1 | 200 (0.92) | 12 (2) | 15.00 | 135 | 2.81 | – | – | – |
| c-fat200.2 | 200 (0.84) | 24 (2) | 26.00 | 35 | 4.93 | 28 | **4** | **1.92** |
| c-fat200.5 | 200 (0.57) | 58 (2) | 77.50 | 21 | 1.71 | 59 | **18** | **0.92** |
| c-fat500.1 | 500 (0.93) | 14 (2) | 17.00 | 543 | 68.08 | 19 | **251** | **31.04** |
| c-fat500.2 | 500 (0.63) | 26 (2) | 30.91 | 485 | 175.32 | 32 | **254** | **109.21** |
| c-fat500.5 | 500 (0.96) | 64 (2) | 74.75 | **89** | 178.01 | 74 | 118 | 116.88 |
| c-fat500.10 | 500 (0.81) | 126 (2) | 155.33 | **13** | 84.70 | 127 | 21 | 31.68 |
| brock200.2 | 200 (0.50) | 12 (7) | 24.49 | **4112** | 105.8 | 36 | 7978 | 93.29 |
| brock200.4 | 200 (0.44) | 17 (10) | 32.83 | **36455** | 393.4 | 50 | 38651 | 200.41 |
| brock400.2 | 400 (0.25) | 29 (20) | 68.19 | 317438 | ≥18000 | 102 | 565318 | ≥18000 |
| brock400.4 | 400 (0.25) | 33 (18) | 67.83 | 261937 | ≥18000 | 102 | 578937 | ≥18000 |
| p_hat300-1 | 300 (0.76) | 8 (7) | 27.00 | **6181** | 755.24 | 29 | 7382 | **395.70** |
| p_hat300-2 | 300 (0.51) | 25 (13) | 45.13 | **10288** | 843.19 | 57 | 12664 | **532.48** |
| p_hat300-3 | 300 (0.26) | 36 (30) | 56.65 | 292443 | 8681.0 | 89 | **286461** | **3740.53** |
| hamming8.4 | 256 (0.36) | 16 (4) | 16.69 | **1** | **0.41** | 33 | 51 | 2.23 |
| keller4 | 171 (0.35) | 11 (7) | 15.30 | 5032 | 28.73 | 26 | **1972** | **12.37** |
| san200.0.7.2 | 200 (0.30) | 18 (13) | 18.00 | **1** | **0.27** | 40 | 158 | 1.52 |
| san200.0.9.1 | 200 (0.10) | 70 (39) | 70.00 | **1** | **0.019** | 89 | 40 | 0.59 |
| san200.0.9.2 | 200 (0.10) | 60 (28) | 60.00 | **1** | **0.033** | 82 | 40 | 0.75 |
| C125.9 | 125 (0.10) | 34 (25) | 43.20 | **3737** | 5.97 | 57 | 4350 | **2.78** |
| C250.9 | 250 (0.10) | [44,48] (35) | 71.91 | 2460234 | ≥18000 | 101 | 4992455 | ≥18000 |
| gen200.9.44 | 200 (0.10) | 44 (32) | 44.00 | **1** | **0.14** | 74 | 255 | 2.21 |
| gen200.9.55 | 200 (0.10) | 55 (27) | 55.62 | **1** | **0.072** | 83 | 40 | 0.80 |
| mann.a27 | 378 (0.01) | 126 (117) | 135.0 | 8272 | 2.24 | 241 | **3191** | **0.89** |
| mann.a45 | 1035 (0.004) | 345 (330) | 360.0 | **56649** | **84.24** | 668 | 124538 | 118.85 |

Table 2: Summary of results on some random and DIMACS benchmark graphs

The results reflect the strength of the new formulation and the effectiveness of the Lagrangian decompo-sition. This new approach outperforms the implementation based on $STAB(G)$ in almost all instances with density at least 30%, as long as the running time is considered. The only significant exception is `hamming8.4`, in which case several subproblems giving the stability number as optimum solution are solved once before being pruned. Naturally, the structure of the graph has great influence on the computation time. One of such relevant structural aspects is the number of maximum stable sets. In cases where this number is small, the decomposition given by the ordering of the vertices gives directly the optimum solution. For instance, for the graphs `c-fat200.1` to `c-fat500.10`, the number of nodes explored in the branch-and-cut trees is smaller when the decomposition is used when compared to the formulation based on $STAB(G)$. For low density graphs, our proposed decomposition is competitive in many cases, but pathological behaviors are observed in some DIMACS benchmark graphs. For instance, in the cases `san200.0.7.2` to `san200.0.9.2`, subproblems are almost as difficult as the original problem based on formulation (2). Another point that can be observed is an acceptable increase in the number of nodes explored, which is the case with several of the graphs `c-fat`, `brock`, and `p_hat`.

| $G$ | Sequential | | 4 cores | | 2×4 cores | |
| --- | --- | --- | --- | --- | --- | --- |
| | Nodes | Time | Work | Speedup | Work | Speedup |
| Random graphs | | | | | | |
| g200.90 | 400.20 | 4.99 | 1 | 1.00 | 1 | 1.00 |
| g200.70 | 1251.2 | 25.76 | 1 | 1.01 | 1 | 1.01 |
| g200.50 | 11922.0 | 64.02 | 1 | 1.19 | 1 | 1.21 |
| g200.30 | 112019.4 | 428.86 | 1 | 1.96 | 1 | 2.31 |
| DIMACS graphs | | | | | | |
| c-fat200.1 | – | – | – | – | – | – |
| c-fat200.2 | 4 | 1.92 | 1 | 1.00 | 1 | 1.00 |
| c-fat200.5 | 18 | 0.92 | 1 | 1.02 | 1 | 1.05 |
| c-fat500.1 | 251 | 31.04 | 1 | 1.00 | 1 | 1.00 |
| c-fat500.2 | 254 | 109.21 | 1 | 1.00 | 1 | 1.00 |
| c-fat500.5 | 118 | 116.88 | 1 | 0.99 | 1 | 0.98 |
| c-fat500.10 | 21 | 31.68 | 2.48 | 1.00 | 2.48 | 1.02 |
| brock200.2 | 7978 | 93.29 | 1 | 1.61 | 1 | 1.62 |
| brock200.4 | 38651 | 200.41 | 1 | 2.65 | 1 | 2.92 |
| p_hat300-1 | 7382 | 395.70 | 1 | 1.12 | 1 | 1.02 |
| p_hat300-2 | 12664 | 532.48 | 1 | 1.00 | 1 | 1.00 |
| p_hat300-3 | 286461 | 3740.53 | 1 | 1.28 | 1 | 1.29 |
| hamming8.4 | 51 | 2.23 | 1 | 2.19 | 1 | 2.42 |
| keller4 | 1972 | 12.37 | 1 | 2.15 | 1 | 2.13 |
| san200.0.7.2 | 158 | 1.52 | 1 | 2.34 | 1 | 2.24 |
| san200.0.9.1 | 40 | 0.59 | 1 | 1.48 | 1 | 2.19 |
| san200.0.9.2 | 40 | 0.75 | 1 | 2.08 | 1 | 2.50 |
| C125.9 | 4350 | 2.78 | 0.99 | 2.96 | 1 | 3.43 |
| gen200.9.44 | 255 | 2.21 | 0.20 | 3.03 | 0.25 | 3.11 |
| gen200.9.55 | 40 | 0.80 | 1 | 1.57 | 1 | 2.29 |
| mann.a27 | 3191 | 0.89 | 1 | 2.23 | 1 | 2.12 |
| mann.a45 | 124538 | 118.85 | 1 | 3.11 | 1 | 3.08 |

Table 3: Parallel performance with 4 and 8 threads

The gain obtained with the multi-threaded parallel implementation is shown in Table 3, for 4 and 8 threads. Two measures are used to evaluate the performance of this parallel implementation. First, the variation of the number of nodes of the branch-and-cut tree explored is given in the column entitled "work". We observe no significant variation in this measure for all graphs, except for `gen200.9.44`, in which case the parallel execution explored much less nodes. Second, the speedup given by sequential time/parallel time is larger for denser graphs. In general, we observe that the Java mechanism used to schedule the threads does not incur any significant overhead. Another fact that should be noticed is that, for the graph `mann.a45`, the parallel implementation of the decomposition performs better than the implementation of $STAB(G)$.

## 5 Concluding remarks

A decomposition for a new formulation for the SSP is proposed and analyzed. An empirical comparison of this decomposition with the standard formulation shows a significant improvement in the efficiency of the branch-and-cut algorithm. According to experimental observations, the standard formulation performs well with sparse graphs. The decomposition leads to a better performance with denser graphs. Additional gain in performance is attained with a parallel implementation that explore the processing power of multicore systems.

As directions for further research, the effectiveness of the Lagrangian decomposition of the representatives formulation (and its shared memory parallel implementation) for the SSP makes this approach very promising for other problems involving stable sets, like the vertex coloring problem and its generalizations.